\documentclass[twocolumn,tighten]{aastex62}
\usepackage{graphicx}
\usepackage{dcolumn}   
\usepackage{amssymb,amsmath,framed,mathtools}

\usepackage{bm}
\expandafter\ifx\csname package@font\endcsname\relax\else
 \expandafter\expandafter
 \expandafter\usepackage
 \expandafter\expandafter
 \expandafter{\csname package@font\endcsname}
\fi
\def\bq{\begin{equation}}
\def\eq{\end{equation}}
\def\bqy{\begin{eqnarray}}
\def\eqy{\end{eqnarray}}

\def\calo{\mathcal{O}}

\begin{document}
\title{Role of planetary obliquity in regulating atmospheric escape: G-dwarf vs. M-dwarf Earth-like exoplanets}

\correspondingauthor{Chuanfei Dong}
\email{dcfy@princeton.edu}

\author{Chuanfei Dong}
\affiliation{Department of Astrophysical Sciences, Princeton University, Princeton, NJ 08544, USA}
\affiliation{Princeton Center for Heliophysics, Princeton Plasma Physics Laboratory, Princeton University, Princeton, NJ 08544, USA}

\author{Zhenguang Huang}
\affiliation{Center for Space Environment Modeling, University of Michigan, Ann Arbor, MI 48109, USA}

\author{Manasvi Lingam}
\affiliation{Institute for Theory and Computation, Harvard University, Cambridge MA 02138, USA}

\begin{abstract}
We present a three-species (H$^+$, O$^+$ and e$^-$) multi-fluid magnetohydrodynamic (MHD) model, endowed with the requisite upper atmospheric chemistry, that is capable of accurately quantifying the magnitude of oxygen ion losses from ``Earth-like'' exoplanets in habitable zones, whose magnetic and rotational axes are roughly coincidental with one another. We apply this model to investigate the role of planetary obliquity in regulating atmospheric losses from a magnetic perspective. For Earth-like exoplanets orbiting solar-type stars, we demonstrate that the dependence of the total atmospheric ion loss rate on the planetary (magnetic) obliquity is relatively weak; the escape rates are found to vary between $2.19 \times 10^{26}$ s$^{-1}$ to $2.37 \times 10^{26}$ s$^{-1}$. In contrast, the obliquity can influence the atmospheric escape rate ($\sim$ $10^{28}$ s$^{-1}$) by more than a factor of $2$ (or $200\%$) in the case of Earth-like exoplanets orbiting late-type M-dwarfs. Thus, our simulations indicate that planetary obliquity may play a weak-to-moderate role insofar as the retention of an atmosphere (necessary for surface habitability) is concerned. \\
\end{abstract}

\section{Introduction} \label{SecIntro}
Over the past decade, much attention has been directed toward understanding what factors contribute to exoplanetary habitability \citep{CBB16}. In particular, it is widely accepted that orbital parameters play a major role in governing habitability \citep{SBJ16}. One of the chief orbital parameters is the obliquity (axial tilt). The fact that Earth's obliquity is subject to only mild fluctuations is believed to play a vital role in maintaining its stable climate. 

In consequence, numerous studies have analyzed how obliquity affects planetary climate \citep{SMS09,FMO14,RCB17,Kang19}. There is broad consensus that climate is sensitive to changes (even minimal ones) in obliquity as it can transition from one stable state to another \citep{WP03,LPL15,KRS17,Colose19}. A number of observational techniques based on amplitude and frequency modulations in light curves arising from thermal emission and scattered light have been proposed for inferring planetary obliquity \citep{GW04,SSH16,Rau17,KT17}. Thermal phase curves of Hot Jupiters have already yielded constraints on planetary obliquity; for example, CoRoT-2 b has an inferred obliquity of $45.8^{\circ} \pm 1.4^{\circ}$ \citep{AML19}.

Exoplanets around M-dwarfs are typically anticipated to have very low (or zero) obliquities due to rapid tidal energy dissipation \citep[e.g.,][]{heller11}. This effect may be particularly pronounced for the inner planets of multi-planet systems such as the Kepler-186 system \citep{BRS14}. Nevertheless, there are mechanisms that permit the existence of high obliquity M-dwarf exoplanets. Perhaps the most famous among them are the ``Cassini states'' that involve the precession of the planet's spin and orbital angular momenta at the same rate \citep{Col66,Pea69,WH05}; the Moon has an non-zero obliquity of $6.7^{\circ}$ due to this reason. Hence, it is feasible for certain M-dwarf exoplanets to have high obliquities \citep{Dom09,WLT16,SL18}; see also \citet{ML19} and \citet{MB19}.

Another factor that plays a vital role in regulating surficial habitability is the presence of an atmosphere. Moreover, an atmosphere also permits the detection of biosignature gases (e.g., molecular oxygen) via spectroscopy \citep{Kal17,SKP18,Mad19}. Recent numerical and theoretical studies indicate that both magnetized and unmagnetized planets around M-dwarfs might be particularly susceptible to the depletion of $\sim 1$ bar planetary atmospheres over sub-Gyr timescales due to the high ultraviolet fluxes and intense stellar winds they experience \citep{GGD17,DLMC,DJL,DLM18,LL18,LL19,aira19}.

In view of the preceding discussion, it is worthwhile asking the question: how does obliquity regulate atmospheric escape rates? It is, however, important to note that the rotational and magnetic axes of Earth are separated only by $\sim 10^\circ$. Hence, it is plausible that these two axes are potentially aligned for Earth-like planets as well, viz., the angle between the two might be fairly small. In other words, the magnetic obliquity, i.e. the angle between the magnetic axis and orbital axis, may approximately coincide with the conventional planetary obliquity. Furthermore, as the magnetic obliquity determines the orientation of the planetary magnetic field, it can influence magnetospheric properties and the resultant escape of atmospheric ions, which forms the subject of this Letter. 

Thus, in this study, we opt to perform a parametric analysis of how magnetic obliquity affects the atmospheric ion loss from magnetized exoplanets. We focus on two distinct examples due to their astrobiological relevance: an Earth-like planet around a solar-type star and an Earth-like planet around a late-type M-dwarf using TRAPPIST-1 as a proxy.

\section{Multi-fluid MHD Model} \label{SecModDes}
A multi-fluid MHD model  is utilized herein to simulate the interaction between stellar winds and exoplanets and the concomitant atmospheric ion loss from Earth-like worlds. On such worlds, note that the major neutral and ion species in planetary upper atmosphere are atomic oxygen and O$^{+}$ respectively \citep{MWD18}, owing to which we focus on O$^{+}$ - the dominant escaping species - in our model. The multi-fluid MHD model we develop and employ herein possesses separate continuity, momentum and pressure equations for each species (see the Appendix for detail). Despite the higher computational expense incurred by the multi-fluid MHD approach, they are more realistic and accurate than traditional single-fluid approach \citep{Toth12,dong14,dong18c}.

To summarize, (\ref{ionmass})-(\ref{Ue}) in the Appendix permit us to simulate the O$^{+}$ ions, stellar wind protons (H$^{+}$) and electrons (e$^{-}$) with complete self-consistency; all essential interactions between fluids as well as their individual evolution are accounted for. More specifically, both elastic and inelastic collisions and the associated heating and cooling terms are explicitly present in our model. It is important to recognize that in addition to photo-electron heating, our model also incorporates \emph{Joule heating}, as seen from manipulating the second term in the square brackets on the RHS of (\ref{elecP}) along the following lines.
\begin{eqnarray}
\sum_{t=\mathrm{O,O^+}}\frac{\rho_{e}\nu_{et}}{m_{e}+m_{t}}\left[\frac{2}{3}m_{t}\left(\mathbf{u_{t}-u_{e}}\right)^{2}\right] \propto \frac{\mathbf{J}^2}{\sigma_e} =  \mathbf{J}\cdot\mathbf{E}
\end{eqnarray}
where $m_t/(m_e+m_t) \approx 1$ in view of the fact that the mass of oxygen is much larger than the electron mass.

A number of chemical reactions such as photoionization, electron impact ionization, charge exchange between neutrals and ions, and recombination are included in the form of source terms delineated in the Appendix. Additional details concerning the reactions are outlined in Table \ref{reacoeff}. For a brief discussion of the utility of multi-fluid MHD models in our Solar system, we refer the reader to Section 2 of \citet{DHL17}.

\begin{table*}
\centering
\caption{The elastic collision rates and chemical reaction rates used in the multi-fluid MHD code. For elastic collisions, $Z_s$, $m_s$, $n_s$ and $T_s$ denote the charge state, mass (in amu), number density (in cm$^{-3}$) and temperature (in K) for a given species. Note that $m_{st}=\frac{m_sm_t}{m_s+m_t}$ and $T_{st}=\frac{m_sT_t+m_tT_s}{m_s+m_t}$ represent the reduced mass and temperature. 
}\label{reacoeff}
\begin{tabular}{llll}
\hline
\hline
\multicolumn{2}{c} {Elastic Collision} & Rate (s$^{-1}$) & Reference \\
\hline
& ion-ion ($\nu_{st}$) & $1.27 \times \frac{Z_s^2 Z_t^2 \sqrt{m_{st}}}{m_s} \frac{n_t}{T_{st}^{3/2}}$ &  \citet{schunk2009} \\
& ion-neutral ($\nu_{sn}$) & $ C_{sn}n_n$ \footnote{Both $\nu_{H^+O}$ and $\nu_{O^+O}$ are for resonant ion-neutral interactions, where $\nu_{H^+O} = 6.61 \times 10^{-11} n_OT_{H^+}^{1/2}(1-0.047\log_{10}T_{H^+})^2$ for $T_r > 300 K$, and $\nu_{O^+O} = 3.67 \times 10^{-11} n_OT_{r}^{1/2}(1-0.064\log_{10}T_{r})^2$ for $T_r > 235 K$. $T_r = (T_{i} + T_n)/2$, where $T_i$ and $T_n$ are the ion and neutral temperatures, respectively. Atomic oxygen number density $n_O$ has units of cm$^{-3}$.} &  \citet{schunk2009} \\
& electron-ion ($\nu_{es}$) & $54.5 \times \frac{n_sZ_s^2}{T_e^{3/2}}$ &  \citet{schunk2009} \\
& ion-electron ($\nu_{se}$) & $1.27 \times \frac{\sqrt{m_e}}{m_s} \frac{n_eZ_s^2}{T_e^{3/2}}$ &  \citet{schunk2009} \\
& electron-neutral ($\nu_{en}$) & $8.9 \times 10^{-11} n_n (1+5.7 \times 10^{-4} T_e)T_e^{1/2}$ &  \citet{schunk2009} \\
\hline
\hline
\multicolumn{2}{c} {Chemical Reaction} & Rate (s$^{-1}$)  \\
\hline
\multicolumn{3}{c} {Primary Photolysis and Particle Impact} \\
\hline
& O + $h\nu$ $\rightarrow$  O$^+$ + $e^{-}$   & see table footnote \footnote{The photoionization rate has been appropriately rescaled based on the extreme ultraviolet (EUV) values at Earth and TRAPPIST-1g. The photoionization rate equals 4.13$\times$ 10$^{-7}$  s$^{-1}$ for Earth, and 32.96 $\times$ 10$^{-7}$  s$^{-1}$ for TRAPPIST-1g.} & \citet{schunk2009} \\
 & e$^-$ + O $\rightarrow$ e$^-$ + O$^+$ + $e^{-}$   & see reference & \citet{cravens87} \\
\hline
\multicolumn{2}{c} {Ion-Neutral Chemistry} & Rate (cm$^{3}$ s$^{-1}$) \\
\hline
&  H$^+$ + O $\rightarrow$  H + O$^+$  & $3.75 \times 10^{-10}$  &  \citet{schunk2009} \\
\hline
\multicolumn{2}{c} {Electron Recombination Chemistry} & Rate (cm$^{3}$s$^{-1}$)  \\
\hline
&  O$^+$  + $e^{-}$ $\rightarrow$  O  & $3.7 \times10^{-12}\left(\frac{250}{T_e}\right)^{0.7}$, &  \citet{schunk2009} \\
& H$^+$  + $e^{-}$ $\rightarrow$  H  & $4.8 \times10^{-12}\left(\frac{250}{T_e}\right)^{0.7}$,  &  \citet{schunk2009} \\
\hline
\hline
\end{tabular}
\end{table*}

\section{Simulation set-up} \label{SecSim}

\begin{figure*}[!ht]
\centering
\includegraphics[width=2\columnwidth]{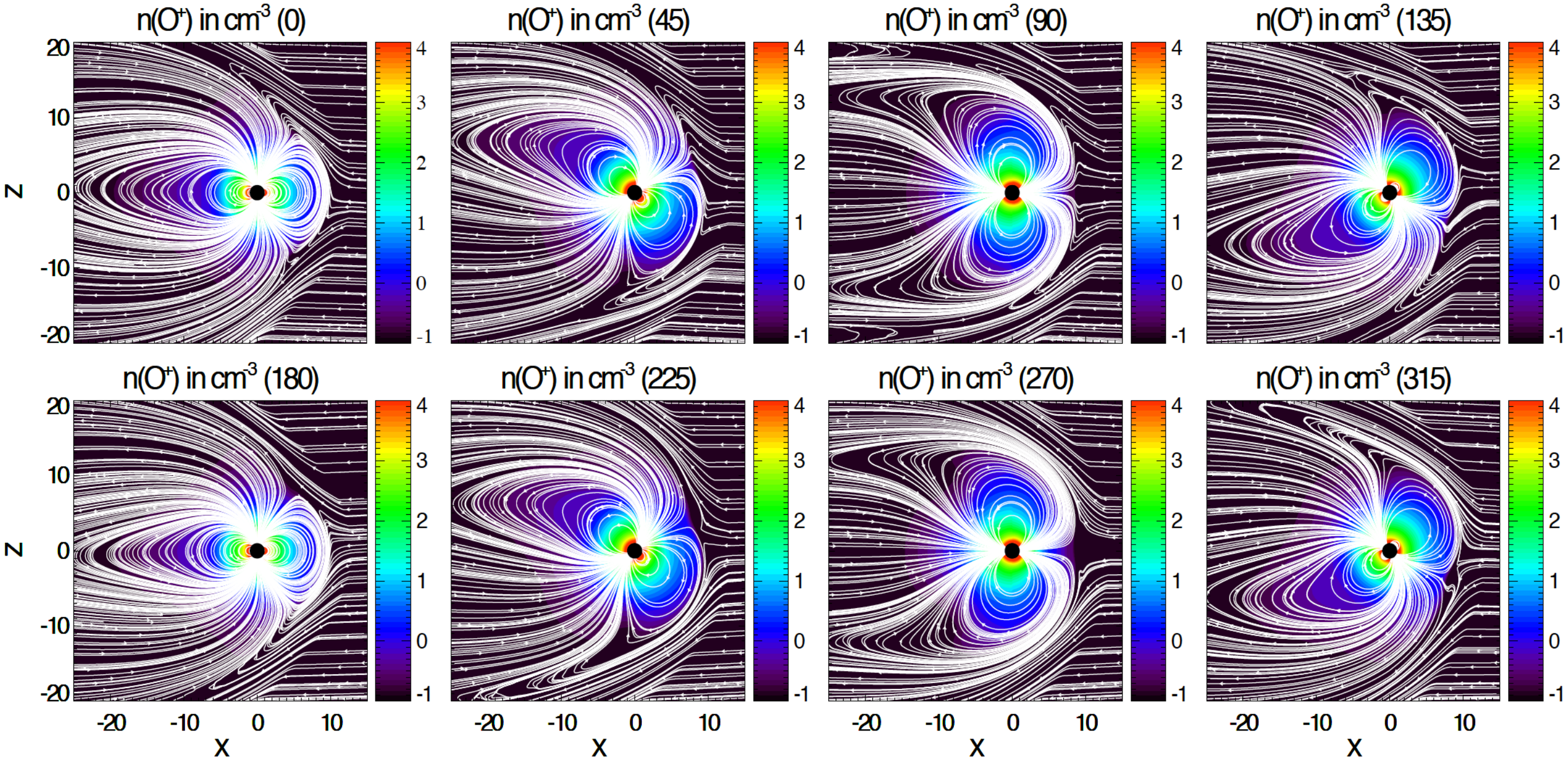}
\includegraphics[width=2\columnwidth]{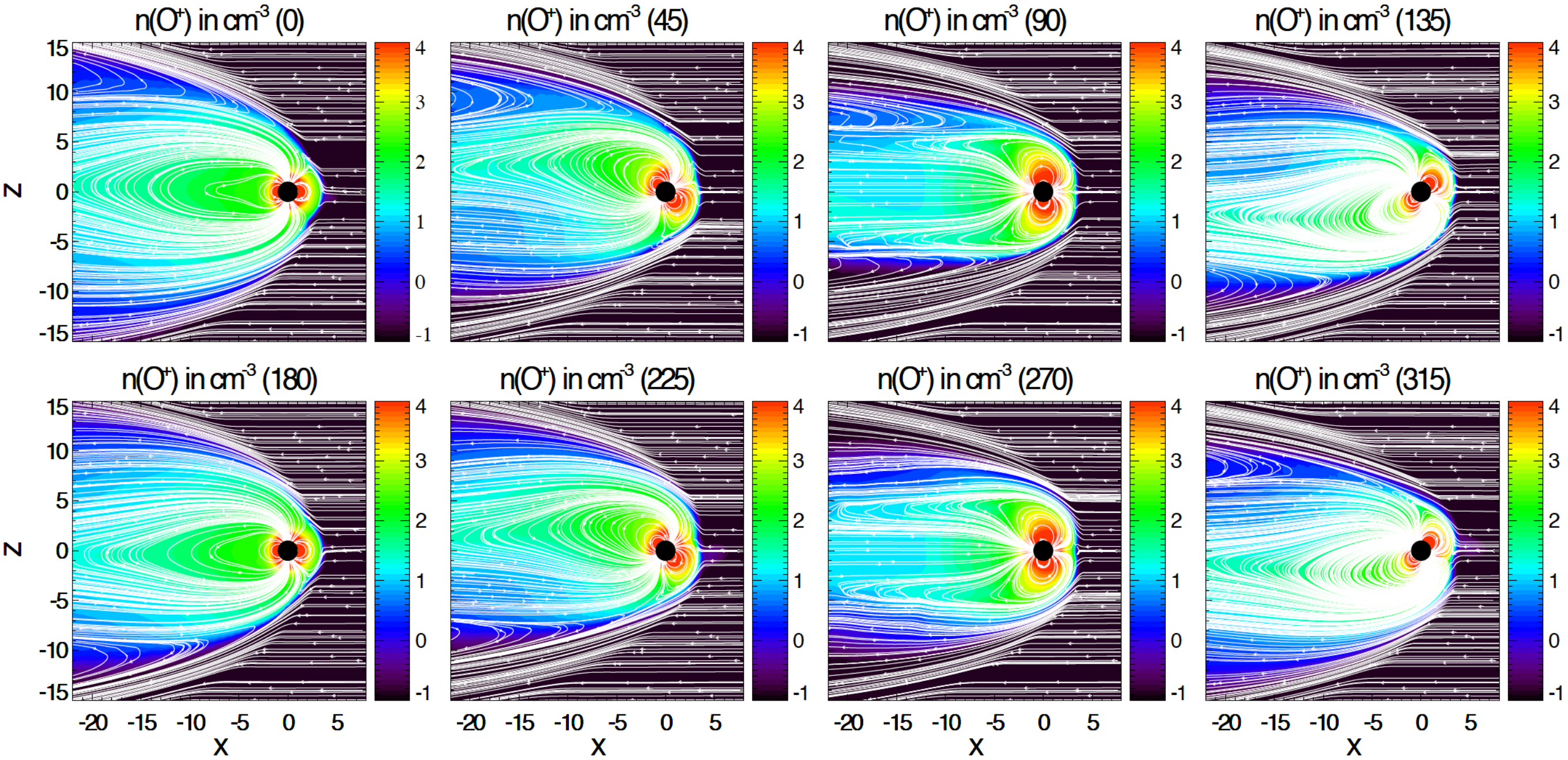}
\caption{Logarithmic scale contour plots of the O$^+$ ion density with magnetic field lines (in white) in the meridional plane based on the stellar wind conditions at current Earth (top two rows) and at TRAPPIST-1g (bottom two rows). Different plots correspond to various choices of the planetary magnetic obliquity. Note the same colorbar range but the different box size between the top two rows and the bottom two rows.}\label{contour}
\end{figure*}

\begin{table*}
\caption{The stellar wind input parameters are based on: (i) typical solar wind parameters at 1 AU in the current epoch \citep{schunk2009}, (ii) stellar wind parameters simulated at TRAPPIST-1g \citep{DJL}. For a fair comparison, we assume that both stellar wind velocities only have an $x$ component, and stellar magnetic fields are located in the x-y plane as the nominal case. Here, the radiation flux received at Earth refers to the solar cycle moderate conditions and the radiation flux received at TRAPPIST-1g is based on estimates provided in \citet{Bol17} and \citet{bourrier17}.} \label{SW}
\centering
\begin{tabular}{llllllll}
\hline
\hline
 & n$_{sw}$ & v$_{sw}$ & IMF  & Radiation & Obliquity & O$^+$ loss rate & Normalization\footnote{The normalization refers to the oxygen ion escape rate normalized to the canonical zero obliquity case.}  \\
 & (cm$^{-3}$) & (km/s) & (nT) & flux & (degree) & (s$^{-1}$) & (\%) \\
\hline
 &  &   & & & 0 & 2.19$\times$10$^{26}$ & 100.0\\ 
 &  &   & & & 45 & 2.21$\times$10$^{26}$ & 101.1 \\
 &  &   & & & 90 & 2.37$\times$10$^{26}$ & 108.2 \\ 
G-dwarf & 8.7  & (-468, 0, 0)  & (-4.4, 4.4, 0) & at Earth & 135 & 2.21$\times$10$^{26}$ & 100.9 \\
planets &  &   & & & 180 & 2.19$\times$10$^{26}$ & 100.0 \\ 
  &  &   & & & 225 & 2.19$\times$10$^{26}$ & 100.3 \\
 &  &   & & & 270 & 2.27$\times$10$^{26}$ & 103.9 \\ 
  &  &   & & & 315 & 2.19$\times$10$^{26}$ &100.3 \\
\hline
 &  &   & & & 0 & 1.01$\times$10$^{28}$ & 100.0 \\
 &  &   & & & 45 & 1.21$\times$10$^{28}$ & 119.2 \\
 &  &   & & & 90 & 2.16$\times$10$^{28}$ & 213.1 \\ 
M-dwarf & 1948.2  &  (-636.7, 0, 0) & (-68.6, 6.2, 0) & at TRAPPIST-1g & 135 & 1.22$\times$10$^{28}$ & 120.4 \\
planets &  &   & & & 180 & 1.01$\times$10$^{28}$ & 99.9 \\ 
  &  &   & & & 225 & 1.13$\times$10$^{28}$ & 111.7 \\
 &  &   & & & 270 & 2.01$\times$10$^{28}$ & 198.1 \\ 
  &  &   & & & 315 & 1.12$\times$10$^{28}$ & 110.8 \\
\hline
\hline
\end{tabular}
\end{table*}

\begin{figure*}[!ht]
\centering
\includegraphics[width=2.0\columnwidth]{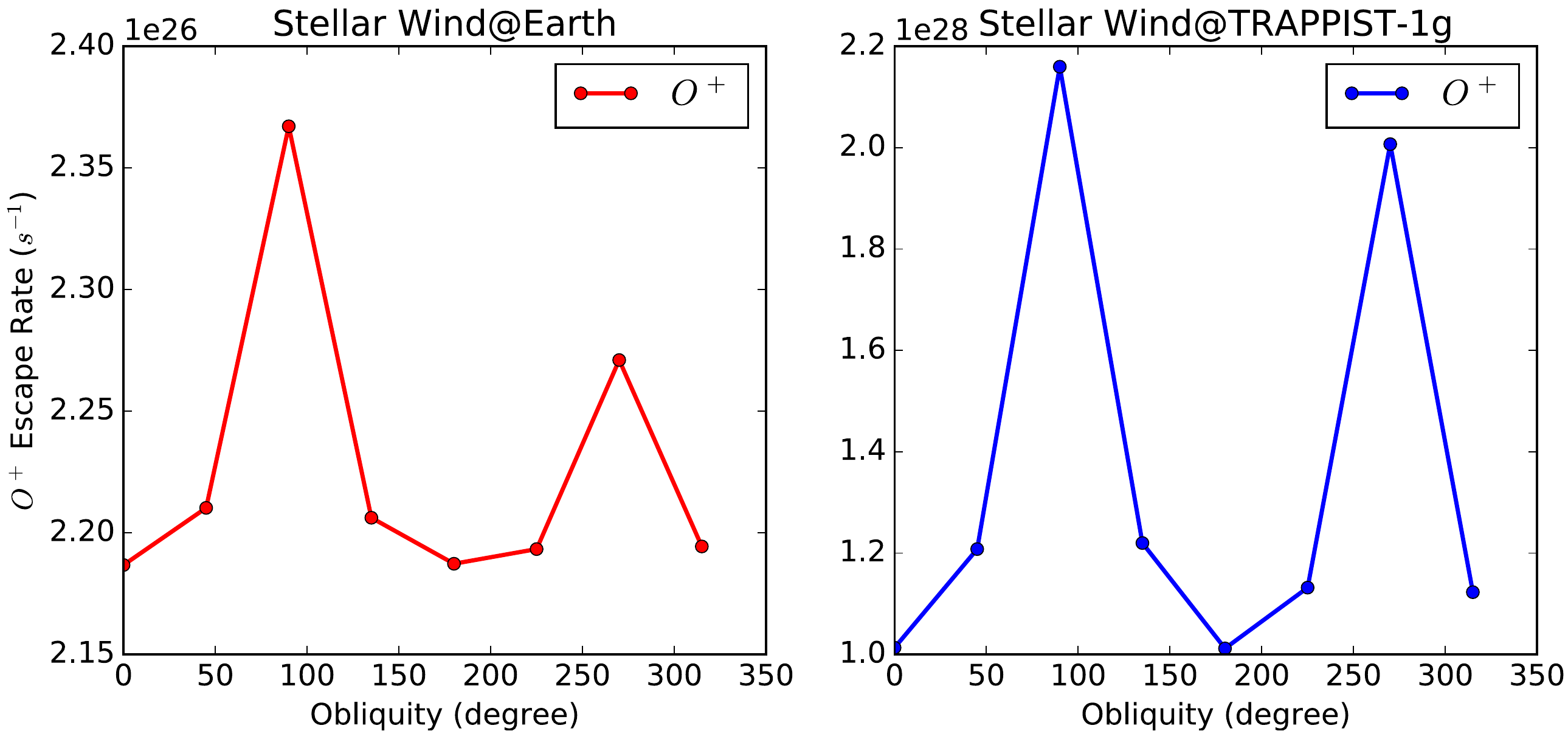}
\caption{Oxygen ion escape rate for different values of the planetary (magnetic) obliquity. Note that the scales of the vertical axis in the two panels are different.}\label{OblAng}
\end{figure*}

As mentioned in Sec. \ref{SecIntro}, we investigate two different Earth-like planets by changing the host star they orbit; this endeavor is important because stellar parameters influence numerous aspects of habitability \citep{LL19}. The first involves a G-type star with the Sun constituting the proxy. The second is based on TRAPPIST-1 because it is a well-known example of late M-dwarfs. 

To investigate Earth-like planets, we make use of Earth's thermospheric temperature profile \citep{schunk2009} and specify a fiducial surface pressure of $1$ bar and magnetic dipole moment equal to the Earth's. It is worth noting that Earth's thermosphere is much hotter ($\sim 1000$ K) than the surface and lower atmosphere. Hence, a surface temperature change of $\sim 20$ K induced by obliquity as per climate models \citep{Kang19} is not likely to affect the upper atmosphere significantly. As noted previously in Sec. \ref{SecModDes}, O$^{+}$ is the primary ion undergoing atmospheric escape in our model because we are interested in exoplanets that resemble Earth. It is feasible to select alternative atmospheres (e.g., resembling Venus), along the lines of \citet{DLMC,DJL}, and study the effects of varying magnetic obliquity but this is left for future publications. 

Our simulation domain commences at $100$ km, where the density of O$^{+}$ ions is roughly in photochemical equilibrium. In our numerical model, float boundary conditions are employed for the velocity $\mathbf{u}$ and the magnetic field $\mathbf{B}$. The simulation box is extended up to $200$ planetary radii by means of a non-uniform spherical grid. In the lower ionosphere and thermosphere, the lowest spatial resolution of $\sim 10$ km - several times smaller than the thermospheric scale height - is attained to capture fine-scale variations in the upper atmosphere. The angular (i.e., horizontal) resolution is 3$^{\circ}$ $\times$ 3$^{\circ}$. The equations in the Appendix are solved by means of an upwind finite-volume scheme \citep{Toth12}; see \citet{DHL17} for further details.

Since we are studying Earth-like planets around solar-type stars and late M-dwarfs, we require the appropriate stellar wind parameters to compute self-consistent ion escape rates. For the Sun, current solar wind parameters are adopted from \citet{schunk2009}. Given that we use TRAPPIST-1 as a proxy, we use the simulated stellar wind parameters from \citet{DJL} at TRAPPIST-1g. The parameters for the two cases are presented in Table \ref{SW}. The Planet-Star-Orbital (PSO) coordinates are used herein. In this system, the X-axis is directed from the planet toward the star, the Z-axis is normal to the planet's orbital plane, and the Y-axis is perpendicular to the X- and Z-axes.

\section{Results} \label{SecRes}

\begin{figure}[!ht]
\includegraphics[width=0.88\columnwidth]{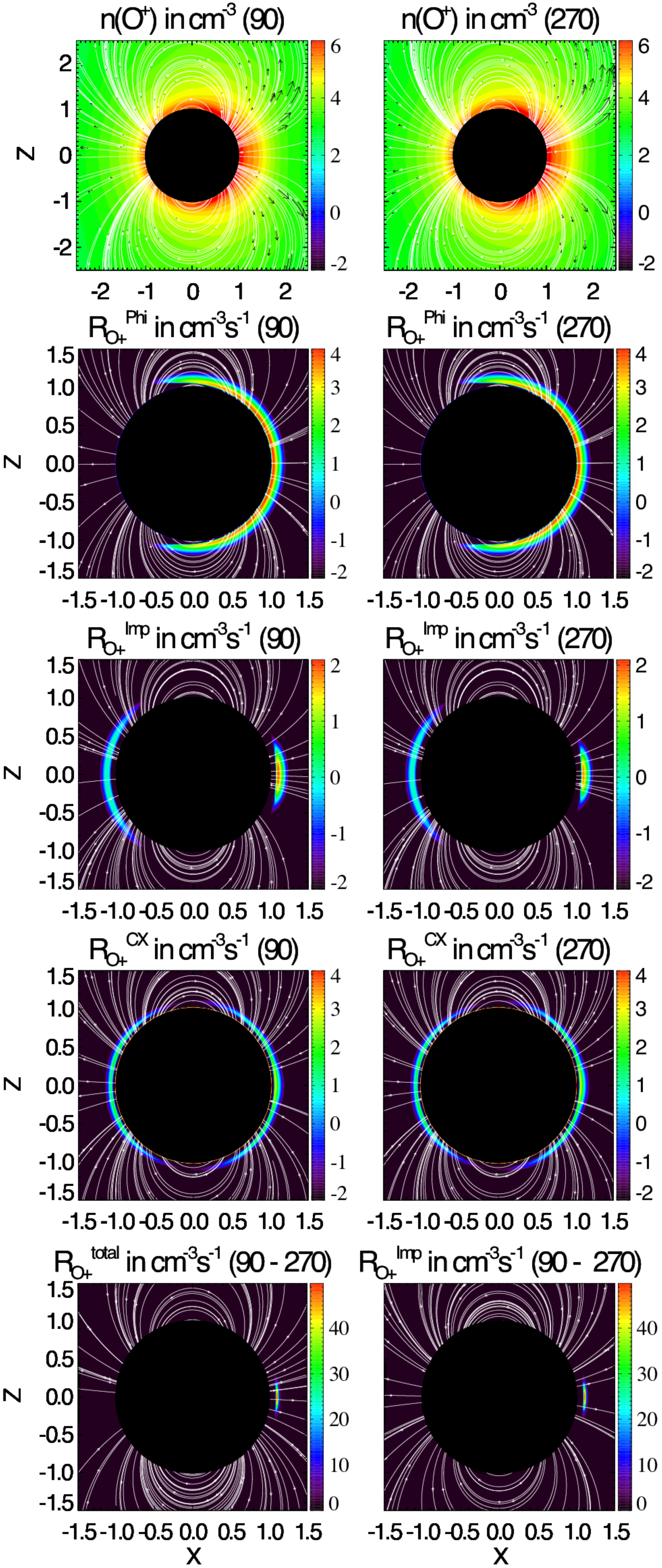}
\caption{\emph{First row:} Logarithmic scale contour plots of the O$^+$ ion density with O$^+$ ion velocity vectors (in black) and magnetic field lines (in white) in the meridional plane for the M-dwarf planet with obliquities of $90^{\circ}$ and $270^{\circ}$, respectively. The black arrows depict both the direction and the magnitude of O$^+$ ion velocities. \emph{Second to fourth rows:} Logarithmic scale contour plots of the photoionization rate R$_{O+}^\mathrm{Phi}$, electron impact ionization rate R$_{O+}^\mathrm{Imp}$, and charge exchange rate R$_{O+}^\mathrm{CX}$ (with stellar wind protons) of O$^+$ ions. \emph{Last row:} The difference in total ionization rate and electron impact ionization rate between $90^{\circ}$ and $270^{\circ}$.}\label{ionization}
\end{figure}

Recall that the our primary intention is to determine how planetary obliquity regulates the atmospheric ion loss from a magnetic perspective. The final results are depicted in Table \ref{SW}. There are several interesting features that stand out in both cases, viz., solar-type stars and late M-dwarfs.

First, the atmospheric ion escape rate is $\calo(10^{26}s^{-1})$ in the case of G-dwarf Earth-like planets, whereas the escape rate increases to $\calo(10^{28}s^{-1})$ for M-dwarf planets due to the extreme stellar wind conditions and high energy radiations in the close-in habitable zones (HZs). In other words, a $\sim 1$ bar atmosphere of an Earth-like planet would take $\calo(10^{10})$ yrs to be depleted for a G-type star and $\calo(10^8)$ yrs for an M-dwarf based on normal stellar wind conditions.   

Second, the variation in the atmospheric ion escape rates is virtually independent of the magnetic obliquity for an Earth-analog around a solar-type star. We find that the total variation is less than $10\%$. In contrast, when we consider an Earth-like planet around a late M-dwarf, we determine that the variation is modest (but non-negligible); in quantitative terms, the maximum escape rate is more than twice (or $200\%$) the minimum value. The chief reason why the obliquity plays a weak role in determining the escape rate for solar-type stars stems from the temperate stellar wind and radiation in HZs. 

As shown in Fig. \ref{contour}, the magnetosphere of the G-dwarf planet is larger than that of the M-dwarf planet; therefore, regardless of magnetic obliquity's value, the ionosphere does not experience much difference. On the other hand, for an Earth-analog around TRAPPIST-1, the dual effect of the compressed magnetosphere and high energy radiation makes the ion sources (e.g., electron impact ionization and charge exchange) more sensitive to the magnetic configuration. 

Third, we see that the maximal ion escape rate is attained at a magnetic obliquity of $90^\circ$ whereas the minimum occurs at $0^\circ$ or $180^\circ$ (Fig. \ref{OblAng}). While the atmospheric escape rates at $0^\circ$ and $180^\circ$ are nearly the same, there is a clear distinction between $90^\circ$ and $270^\circ$. The reason behind the latter behavior has to do with the relative orientation of the interplanetary magnetic field (IMF) and the planetary magnetic field. At the magnetic obliquity of $90^\circ$, the IMF can directly connect to the dayside planetary surface due to the field polarity, whereas the IMF can only connect to the nightside surface at the magnetic obliquity of $270^\circ$; see the third column of Fig. \ref{contour}. Therefore, stellar wind particles, especially electrons (with relatively low energy) can be transported along the field lines and ionize the atomic oxygen in the upper atmosphere via impacts as shown in Fig. \ref{ionization}. 

Fig. \ref{contour} depicts the contour plots of the O$^+$ ion density for an Earth-like planet around a solar-type star (top two panels) and a late M-dwarf (bottom two panels) at different values of the magnetic obliquity. Consistent with the results in Table \ref{SW}, more O$^+$ ions escape from the M-dwarf planet due to the extreme stellar wind and radiations. Fig. \ref{OblAng} shows the corresponding atmospheric oxygen ion escape rates, where two peaks clearly occur at magnetic obliquities of $90^\circ$ and $270^\circ$; the former is slightly higher than the latter.

In the $90^\circ$ or $270^\circ$ cases, the cusp region (i.e., the region filled with open magnetic field lines) directly faces the star and thus the stellar wind. Therefore, the stellar wind particles can impact the dayside upper atmosphere relatively easily and deposit their energy. Compared to the $270^\circ$ case, the $90^\circ$ case possesses a better magnetic connection with the planet (based on our model setup), thereby achieving a slightly higher escape rate primarily due to the high electron impact ionization.

In order to demonstrate this point, the oxygen ion density and velocity (first row) and different ionization rates (second to fourth rows) are illustrated for an Earth-like planet around a late M-dwarf in Fig. \ref{ionization}. The ion outflow of O$^+$ (namely, the polar wind) driven primarily by the electron pressure gradient, $\nabla p_e$, is rendered in the upper panel of Fig. \ref{ionization}. The bottom panel of Fig. \ref{ionization} shows the difference of the total ionization rate between $90^\circ$ and $270^\circ$ (left column), resembling the difference in electron impact ionization rate (right column); in other words, the difference in total ionization rate is mainly regulated by electron impact ionization.

Lastly, over an extended period of time, due to the polarity reversals of stellar and planetary magnetic fields \citep{Gla13}, variations in the escape rate with obliquity may get averaged out.

\section{Conclusions}
In this Letter, we have described a sophisticated numerical code for simulating the escape of atmospheric ions from Earth-like (exo)planets. We have incorporated the appropriate upper atmospheric chemistry and evolve each species separately via the multi-fluid MHD equations of Sec. \ref{SecModDes}. This model was applied to study two different exoplanets: the first around a solar-type star and the second orbiting a late M-dwarf, for which TRAPPIST-1 was used as a proxy. Our goal was to determine how the ion escape rates vary with the planetary obliquity from a magnetic perspective.

We found that the maximum escape rates arose at obliquities of $90^\circ$ or $270^\circ$ (depending on field polarities), whereas the minimum rates were attained at $0^\circ$ or $180^\circ$. The reason is that the cusp (comprising open field lines) faces the stellar wind at obliquities of $90^\circ$ or $270^\circ$, and allows the stellar wind particles to deposit their energy in the planetary upper atmosphere. For Earth-like planets around solar-type stars, it is found that the escape rate is virtually independent of the obliquity. On the other hand, for late M-dwarfs, we determined that the escape rate varies by more than a factor of $\sim 2$. 

From our simulations, we found that the timescale required to deplete a $\sim 1$ bar Earth-like atmosphere is $\calo(10^{10})$ and $\calo(10^{8})$ yrs, for solar-type stars and late M-dwarfs, respectively. If we assume that the source of atmospheric O is water from oceans, we find that the mass of Earth's oceans ($M_{oc}$) cannot be depleted over the main-sequence lifetime of a solar-type star. In contrast, for a late M-dwarf we determine that $M_{oc}$ could be depleted over a timescale of $\calo(10^{10})$ yrs, which is shorter than the star's lifetime.

There are two conclusions to be drawn from this finding. When studying Earth-like planets around solar-type stars, at least insofar as the atmospheric ion escape rates are concerned, the effects of obliquity is ostensibly minimal. In contrast, when it comes to exoplanets around late M-dwarfs, obliquity might play an important role. There are, however, two different cases to consider for M-dwarf exoplanets. In the first case, if the atmosphere is completely depleted over a timescale that is orders of magnitude smaller than $1$ Gyr, changing this value by a factor of $\sim 2$ will probably not have major implications for the origin and evolution of life. 

However, let us consider the second case wherein the M-dwarf exoplanet under question has a sufficiently massive atmosphere or rapid outgassing to permit the retention of an atmosphere over a few Gyr. The habitable timescale for biological evolution will be roughly halved as one moves from an obliquity of $0^\circ$ to $90^\circ$. It is therefore instructive to carry out a thought experiment. Suppose that an Earth-like planet can retain an atmosphere for $4$ Gyr for an obliquity of $0^\circ$ and that biological evolution unfolds in a similar fashion as on Earth;\footnote{This is clearly an idealization, but one we adopt to carry out the thought experiment to fruition.} at an obliquity of $90^\circ$, the depletion timescale is $1.9$ Gyr. For this specific hypothetical planet, at $0^\circ$ obliquity, enough time might exist for the emergence of complex multicellular organisms, whereas an obliquity of $90^\circ$ may not suffice for the evolution of eukaryotic analogs and complex multicellularity \citep{Knoll15}.

Some caveats regarding our treatment are worth emphasizing here. We have chosen to vary only the stellar parameters, thus leaving planetary parameters (e.g., size and magnetic field strength) fixed. In actuality, habitable exoplanets are probably very diverse and the escape rates will change accordingly; however, it is plausible that the qualitative trends described herein are still partly valid. Moreover, we have incorporated only O$^{+}$ as it constitutes the dominant ion species in the Earth's ionosphere, but subsequent treatments should incorporate additional minor species. As we utilize a multi-fluid MHD model, kinetic effects contributing to atmospheric ion escape are not included in our model \citep[e.g., see][]{DrStrange}. Finally, the issue of atmospheric depletion is difficult to address comprehensively because it also necessitates knowledge of other pertinent issues including outgassing, bolide impacts, (super)flares and associated phenomena (e.g., coronal mass ejections).

To summarize, planetary magnetic obliquity does not appear to affect atmospheric ion escape rates for habitable planets around solar-type stars, whereas it has a weak-to-moderate influence on the escape rates for late M-dwarf exoplanets. 

\acknowledgments
The authors acknowledge fruitful discussions with Yutong Shan, Anthony Del Genio, Michael Way, Fei Dai, Gongjie Li, and Joshua Winn. CD was supported by NASA grant 80NSSC18K0288. ML was supported by the Institute for Theory and Computation at Harvard University. Resources for this work were provided by the NASA High-End Computing (HEC) Program through the NASA Advanced Supercomputing (NAS) Division at Ames Research Center. The Space Weather Modeling Framework that contains the BATS-R-US code used in this study is publicly available from \url{http://csem.engin.umich.edu/tools/swmf}. For distribution of the model results used in this study, please contact the corresponding author.

\appendix
In this section, we describe the multi-fluid MHD model, endowed with the electron pressure equation, that is used to simulate the oxygen ion loss from Earth-like exoplanets.

The multi-fluid MHD model comprises three fluids, of which two of them are ionic O$^{+}$ and H$^{+}$ (with subscript $s$) and the last is the electron fluid with subscript $e$. For the background neutral species, we employ the subscript $n$. In  the multi-fluid MHD equations, $\rho$, $\mathbf{u}$, $p$, $\overleftrightarrow{I\,}$, $k_B$ and $ \gamma=5/3$ represent the mass density, velocity vector, pressure, identity matrix, Boltzmann constant and specific heat ratio. An extended description of the multi-fluid MHD equations can be found in \citet{rubin14b}, \citet{huang16a}, and \citet{DHL17}:
\begin{eqnarray} \label{ionmass}
&& \frac{\partial \rho_{s}}{\partial t}+\nabla\cdot(\rho_{s}\mathbf{u_{s}})=\mathcal{S}_{s}-\mathcal{L}_{s} \\ \nonumber \label{ionmom}
&&\frac{\partial \left(\rho_{s}\mathbf{u_{s}}\right)}{\partial t}+\nabla\cdot\left(\rho_{s}\mathbf{u_{s}u_{s}}+p_{s}\overleftrightarrow{I\,}\right)=n_{s}q_{s}\left(\mathbf{u_{s}-u_{+}}\right)\times\mathbf{B} \\ \nonumber
&& \hspace{0.5 in} +\frac{q_{s}n_{s}}{en_{e}}\left(\mathbf{J}\times \mathbf{B}-\nabla\mathit{p_{e}}\right)+\rho_s\mathbf{G}\\ 
&& \hspace{0.5 in} +\rho_{s}\sum_{t=\mathrm{all}}\nu_{st}(\mathbf{u_{t}-u_{s}})+\mathcal{S}_{s} \mathbf{u_{n}}-\mathcal{L}_{s} \mathbf{u_{s}} \\  \nonumber \label{ionpress} 
\end{eqnarray}
\begin{eqnarray} \nonumber 
&&\frac{\partial p_{s}}{\partial t}+\left(\mathbf{u_{s}}\cdot\nabla\right)p_{s}=-\gamma_{s} p_{s}\left(\nabla\cdot\mathbf{u_{s}}\right) \\ \nonumber
&&\hspace{0.4 in} + \sum_{t=\mathrm{all}}\frac{\rho_{s}\nu_{st}}{m_{s}+m_{t}}\left[2k_B\left(T_{t}-T_{s}\right)+\frac{2}{3}m_{t}\left(\mathbf{u_{t}-u_{s}}\right)^{2}\right] \\ 
&& \hspace{0.4 in} + k_B\frac{\mathcal{S}_{s}T_{n}-\mathcal{L}_{s}T_{s}}{m_{s}}+\frac{1}{3}S_{s}\left(\mathbf{u_{n}-u_{s}}\right)^{2} \\ 
\label{elecP} \nonumber 
&& \frac{\partial p_{e}}{\partial t}+\left(\mathbf{u_{e}} \cdot \nabla\right)p_{e}= -\gamma_{e} p_{e}\left(\nabla\cdot\mathbf{u_{e}}\right) \\ \nonumber
&& \, +  \sum_{t=\mathrm{s,n}}\frac{\rho_{e}\nu_{et}}{m_{e}+m_{t}}\left[2k_B\left(T_{t}-T_{e}\right)+\frac{2}{3}m_{t}\left(\mathbf{u_{t}-u_{e}}\right)^{2}\right] \\ 
\nonumber 
&& \, - k_B\frac{\mathcal{L}_{e}T_{e}}{m_{e}} + \frac{2}{3}n_{n}(\nu_{ph,n}\mathcal{E}_{ns}^{exc}-\nu_{imp,n}\mathcal{E}_{ns}^{pot})  \\ 
&& \, - \frac{2}{3}n_e n_n\mathcal{R}_{en}^{inelastic} +\frac{1}{3}S_{e}\left(\mathbf{u_{n}-u_{e}}\right)^{2}\\ 
\label{Bind}
&&\frac{\mathbf{\partial B}}{\partial t}=\nabla\times(\mathbf{u_{+}}\times\mathbf{B}-\eta\mathbf{J}) 
\end{eqnarray}
where $\nu$ signifies the collision frequency between two species and $\mathbf{u_+}$ refers to the charge-averaged velocity,
\begin{equation}
\mathbf{u_{+}}=\sum_{s=ions}\frac{q_{s}n_{s}u_{s}}{en_{e}}
\end{equation}
while $\eta$ is the magnetic diffusivity,
\begin{equation}
\eta = \frac{1}{\mu_0\sigma_e} =  \frac{1}{\mu_0}\left(\frac{1}{\sigma_{en}}+\frac{1}{\sigma_{ei}} \right)
\end{equation}
where $\sigma_e$ is the electron conductivity and is composed of both electron-neutral ($\sigma_{en}=e^2n_e/\Sigma_{n'}\nu_{en'}m_e$) and electron-ion ($\sigma_{ei}=e^2n_e/\Sigma_{s'}\nu_{es'}m_e$) collisions. 

The preceding set of equations consists of source ($\mathcal{S}$) and loss ($\mathcal{L}$) terms:
\begin{eqnarray}\label{Ss}
\mathcal{S}_{s}&=&m_{s}n_{s^{\prime}}\left(\nu_{ph,s^{\prime}}+\nu_{imp,s^{\prime}}+\underset{i=\mathrm{ions}}{\sum}k_{is^{\prime}}n_{i}\right) \\
\mathcal{L}_{s}&=&m_{s}n_{s}\left(\alpha_{R,s}n_{e}+\underset{t^{\prime}=\mathrm{neutrals}}{\sum}k_{st^{\prime}}n_{t\prime}\right) \\
\mathcal{S}_{e}&=&m_{e}\underset{s'}{\sum}\,n_{s'}\,(\nu_{ph,s^{\prime}}+\nu_{imp,s^{\prime}}) \\
\mathcal{L}_{e}&=&m_{e}n_{e}\underset{s=\mathrm{ions}}{\sum}\alpha_{R,s}n_{s}.
\end{eqnarray}
Note that the source and loss terms consist of photoionization ($\nu_{ph,s^{\prime}}$), charge exchange ($k_{is^{\prime}}$), electron impact ionization ($\nu_{imp,s^{\prime}}$), and recombination ($\alpha_{R,s}$). 

In addition, inelastic collisions between electrons and O (neutral oxygen) are effective at cooling the former in the lower thermosphere, where collisions occur regularly. Therefore, we incorporate the cooling rate coefficient $\mathcal{R}_{en}^{inelastic}$ (in eV cm$^3$s$^{-1}$) in (\ref{elecP}) by adopting the formalism in \citet{schunk2009}:
\begin{eqnarray} \label{Qinel}
\nonumber
\mathcal{R}_{en}^{inelastic} &=& D^{-1}\{S_{10}\left(1-\exp\left[98.9\left(T_e^{-1}-T_n^{-1}\right)\right]\right)  \\ \nonumber
&& + S_{20}(1-\exp\left[326.6\left(T_e^{-1}-T_n^{-1}\right)]\right) \\ 
&& + S_{21}(1-\exp\left[227.7\left(T_e^{-1}-T_n^{-1}\right)]\right) \},
\end{eqnarray}
where $D=5+\exp(-326.6 T_n^{-1}) + 3\exp(-227.7 T_n^{-1})$, $S_{21} = 1.863 \cdot 10^{-11}$, $S_{20} = 1.191 \cdot 10^{-11}$, and $S_{10} = 8.249 \cdot 10^{-16} T_e^{0.6} \exp(-227.7 T_n^{-1})$.

The optical depth of the neutral atmosphere is determined by employing the numerical formula in \citet{smith1972} to study photoionization. Photoelectrons gain excess energy $\mathcal{E}_{ns}^{exc}$ during the photoionization process and they lose the ionization energy of neutral oxygen, $\mathcal{E}_{ns}^{pot}$, during electron impact ionization \citep{schunk2009}, as seen from Eq. (\ref{elecP}). The number density and velocity of electrons is easy to calculate after imposing quasineutrality and expressing it in terms of the current \citep{Toth12}, 
\begin{eqnarray}
n_e = \frac{1}{e}\sum_{s=\mathrm{ions}}n_sq_s,
\end{eqnarray}
\begin{equation} \label{Ue}
\mathbf{u}_e = \mathbf{u_+} - \frac{{\mathbf{J}}}{e n_e} = \mathbf{u_+} - \frac{{\mathbf{\nabla\times B}}}{\mu_0 e n_e}
\end{equation}
with the last equation following from Amp\'{e}re's law.


\end{document}